\documentclass[useAMS,usenatbib,usegraphicx]{mn2e}

\def\ltsima{$\; \buildrel < \over \sim \;$}
\def\prosima{$\; \buildrel \propto \over \sim \;$}
\def\gsim{\lower.7ex\hbox{\gtsima}}
\def\lsim{\lower.7ex\hbox{\ltsima}}
\def\simgt{\lower.7ex\hbox{\gtsima}}
\def\simlt{\lower.7ex\hbox{\ltsima}}
\def\simpr{\lower.7ex\hbox{\prosima}}
\def\la{\lsim}
\def\ga{\gsim}
\def\lta{\la}

\title[Gas-Phase Metallicity of Centrals and Satellites]
      {The Gas-Phase Metallicity of Central and Satellite Galaxies 
       in the SDSS}
\author[A. Pasquali et al.]
       {Anna Pasquali$^{1}$\thanks{E-mail:pasquali@ari.uni-heidelberg.de}, 
        Anna Gallazzi$^{2}$ and 
        Frank C. van den Bosch$^{3}$\\
$^{1}$Astronomisches Rechen-Institut, Zentrum f\"ur Astronomie der Universit\"at Heidelberg, 
M\"onchhofstrasse 12 - 14, 69120 Heidelberg, Germany\\
$^{2}$Dark Cosmology Centre, Niels Bohr Institute, University of Copenhagen, Juliane Maries
Vej 30, 2100 Copenhagen, Denmark\\
$^{3}$Department of Astronomy, Yale University, P.O. Box 208101, New Haven, CT 06520-8101, USA}

\begin{document}

\date{}

\pagerange{\pageref{firstpage}--\pageref{lastpage}} \pubyear{2002}

\maketitle

\label{firstpage}

\begin{abstract}
  We exploit the galaxy groups catalogue of Yang et al. and the
  galaxy properties measured in the SDSS Data Releases 4 and 7 to
  study how the gas-phase metallicities of star-forming galaxies
    depend on environment. We find that satellite and central
  galaxies follow a qualitatively similar stellar mass ($M_{\star}$) -
  gas-phase metallicity relation, whereby their gas-phase metallicity
  increases with $M_{\star}$.  Satellites, though, have higher
    gas-phase metallicities than equally massive centrals, and this
    difference increases with decreasing stellar mass. We find
  a maximum offset of 0.06 dex at log($M_{\star}/h^{-2}$M$_{\odot})
  \simeq$ 8.25.  At fixed halo mass, centrals are more metal
    rich than satellites by $\sim$0.5 dex on average. This is
    simply due to the fact that, by definition, centrals are the most
    massive galaxies in their groups, and the fact that gas-phase
    metallicity increases with stellar mass. More interestingly, we
    also find that the gas-phase metallicity of satellites increases
    with halo mass ($M_{\rm h}$) at fixed stellar mass. This
  increment is more pronounced for less massive galaxies, and, at
  $M_{\star} \simeq 10^{9} h^{-2} M_{\odot}$, corresponds to $\sim
  0.15$ dex across the range 11 $<$ log($M_{\rm h}/h^{-1}$M$_{\odot})
  <$ 14.  We also show that low mass satellite galaxies have
    higher gas-phase metallicities than central galaxies of the same
    stellar metallicity. This difference becomes negligible for more
    massive galaxies of roughly solar metallicity. We demonstrate that
    the observed differences in gas-phase metallicity between centrals
    and satellites at fixed $M_{\star}$ are not a consequence of
    stellar mass stripping (advocated by Pasquali et al. in order
    to explain similar differences but in stellar metallicity), nor
    to the past star formation history of these galaxies as quantified
    by their surface mass density or gas mass fraction. Rather, we
    argue that these trends probably originate from a combination of
    three environmental effects: (i) strangulation, which prevents
    satellite galaxies from accreting new, low metallicity gas which
    would otherwise dilute their ISM, (ii) ram-pressure stripping of
    the outer gas disk, thereby inhibiting radial inflows of
    low-metallicity gas, and (iii) external pressure provided
    by the hot gas of the host halo which prevents metal-enriched outflows
    from escaping the galaxies. Each of these three mechanisms
    naturally explains why the difference in gas-phase metallicity
    between centrals and satellites increases with decreasing stellar
    mass and with increasing host halo mass, at least
    qualitatively. However, more detailed simulations and observations
    are required in order to discriminate between these mechanisms,
    and to test, in detail, whether they are consistent with the data.
\end{abstract}

\begin{keywords}
galaxies: abundances -- 
galaxies: evolution -- 
galaxies: fundamental parameters --
galaxies: groups: general -- 
galaxies: star formation
\end{keywords}

\section{Introduction}

Star forming galaxies in the local Universe define a fundamental
plane, where their star formation rate (SFR) and gas-phase metallicity
(typically 12 $+$ log(O/H)) correlate with their stellar mass
($M_{\star}$, cf. Lara-L\'opez et al. 2010; Mannucci et al. 2010). The
individual relations in this plane have been long known; for example,
\citet{b3} were the first to recognize the existence of a correlation
between galaxy magnitude and gas-phase metallicity, whereby more
luminous galaxies exhibit higher metallicities.  More observational
evidence in support of such a dependence was later provided by Garnett
\& Shields (1987), Skillman et al. (1989) and Zaritsky et al. (1994)
among others. The advent of the Sloan Digital Sky Survey (SDSS)
made it possible for \citet{b7} to transform the galaxy luminosity -
metallicity relation into the more fundamental dependence of gas-phase
metallicity on galaxy stellar mass, where the interstellar medium
(ISM) of more massive galaxies is metal richer. Using the SDSS,
\citet{b8} and \citet{b10} established a correlation between the SFR
of galaxies and their stellar mass, whereby more massive, star forming
galaxies sustain larger star formation rates (see also Ellison et
al. 2008). The star formation rate has been seen to
correlate also with the gas-phase metallicity of galaxies, although in
a way that differs according to the galaxy stellar mass. As pointed
out by \citet{b2}, the gas-phase metallicity of low--mass galaxies
decreases as their SFR increases, while high--mass galaxies do not
show any significant dependence of their gas-phase metallicity on SFR
(see also Yates et al. 2012).

Such an interplay among $M_{\star}$, SFR and gas-phase metallicity of
galaxies is usually explained in terms of stellar feedback triggering
gas and metals outflows (see Edmunds 1990, Lehnert \& Heckman 1996;
Frye, Broadhurst \& Ben\'itez 2002; Garnett 2002; Tremonti et
al. 2004; Kobayashi, Springel \& White 2007; Scannapieco et al. 2008;
Weiner et al. 2009; Spitoni et al. 2010; McCarthy et al. 2011; Peeples
\& Shankar 2011).  \citet{b2} envisaged a condition of steady--state
in the local Universe, where infall of metal-poor gas (diluting the
galaxy gas-phase metallicity and sustaining its star formation
activity) occurs together with outflow of metal-rich gas whose
efficiency should depend on the galaxy mass and SFR.  Alternative
explanations of the fundamental plane of star-forming galaxies invoke
downsizing (where the star formation efficiency depends on galaxy
mass, e.g. Brooks et al. 2007; Mouhcine et al. 2008; Calura et
al. 2009), a dependence of the initial mass function on galaxy mass
(K\"oppen, Weidner \& Kroupa 2007) or a model of pure infall of
metal-poor gas (Finlator \& Dav\'e 2008; Dav\'e et al. 2010).

One can easily expect that environment may intervene in shaping the
fundamental plane of star-forming galaxies. Environmental processes,
such as strangulation (i.e. the removal of the gas reservoir of satellites 
after they are accreted by their host halo), ram pressure stripping (i.e. the removal
of gas from satellites moving through a dense intracluster medium), galaxy
harassement (i.e. high-speed impulsive galaxy encounters) and galactic wind
confinement (i.e. due to the pressure exerted by the hot gaseous atmosphere
of the host halo, which can inhibit galactic winds), are believed to alter the 
inflows/outflows of gas experienced by galaxies, modifying their star formation
activity and their gas-phase metallicities with
respect to that predicted by closed-box evolution.

Observations of the ISM in galaxies residing in the Lynx-Cancer void show that the
gas-phase metallicity of these galaxies is lower by 30$\%$ on average
than that of equally-massive galaxies in higher-density environments
(Pustilnik, Tepliakova \& Kniazev 2011). 

Nearby clusters (such as Coma, Hercules and Virgo) have been
extensively observed in search of environmental effects on the
properties of star-forming galaxies. For example, \citet{b30} and
\citet{b31} found that the ISM of spiral galaxies in Virgo is on
average metal-richer than that of comparable field
galaxies. When a distinction is made on the basis of their HI content,
the Virgo spirals that are HI deficient and also located closer to the
cluster core are seen to have a higher gas-phase metallicity and a
significantly lower birthrate (i.e. the ratio of newly-born stars to
the stars formed in the past; Gavazzi et al. 2002) than those with
normal HI content, which are usually located in the cluster outskirts
and have similar metallicity as field galaxies. More recently,
\citet{b32} have shown that spiral galaxies in the Hercules cluster
are chemically evolved with an oxygen abundance close to solar,
and their gas-phase
metallicity does not depend on the local density within the
cluster. Interestingly, their H$\alpha$ emission is less spatially
extended than their optical disks and their star formation activity
takes place preferentially in their inner regions. This result
supports the picture where ram-pressure stripping due to the
intergalactic medium (IGM) has truncated the HI disk of these spirals,
with the result of preventing inflows of metal-poor gas from the outer
disk and halo which would otherwise dilute the gas-phase
metallicity in the inner regions of these galaxies. Suppression of
infall of metal-poor gas due to environment would also be an explanation 
for the higher gas-phase metallicity of cluster spirals.

\begin{table*}
\centering
\caption{Samples statistics}
\begin{tabular}{llrrc}
\hline
            & Description & Centrals & Satellites & Satellite Fraction\\
\hline
Sample $S$  & galaxies with spectral S/N $\geq$ 20 and measured age and Z &  70,067 & 13,626 
& 19.4$\%$\\
Sample $G$  & galaxies with EW (H$\beta) \leq$ -3 \AA\/ and measured 12 + log(O/H) &  69,688 
& 14,182 & 20.4$\%$\\
Sample $C$  & galaxies that are both in Samples $S$ and $G$ &   8,134 &  1,595 & 19.6$\%$\\
\hline
\end{tabular}
\end{table*}

As for dwarf galaxies, the studies of \citet{b33} and \citet{b34} did
not reveal a significant dependence of gas-phase metallicity on
environment.  On average, equally-bright dwarfs residing in the field
and in Virgo do not differ in their O/H abundance, but some Virgo
dwarfs can show a gas deficiency when compared with equally metal-rich
dwarfs in the field. Such a gas deficiency seems to correlate with the
X-ray surface brightness of the IGM, thus suggesting that ram-pressure
stripping due to the IGM deprived these Virgo dwarfs of some of their
gas. Ram-pressure stripping has been observed `in action' for already
a number of galaxies in Virgo (e.g. Crowl et al. 2010), whose HI disks
appear distorted and truncated with respect to their stellar disks. A
somewhat stronger dependence on environment was established by
\citet{b32} for the dwarf galaxies in the Hercules cluster, whose
gas-phase metallicity is seen increasing with local
density. Metal-poorer dwarfs are associated with lower densities
(typically the infall regions of the cluster) and with a stronger
H$\alpha$ emission with respect to the global dwarf population in the
cluster (Mahajan, Haines \& Raychaudhury 2010), a sign that they
undergo a starburst phase upon accretion onto the cluster.  Such a
star formation activity has to get quenched at later times in order to
reproduce the observed fractions of red-and-dead satellites (e.g. van
den Bosch et al. 2008; Kang \& van den Bosch 2008; Kimm et
al. 2009; Pasquali et al. 2009; Weinmann et al. 2009).

The large statistical power of the SDSS survey has not yet provided a
definitive assessment of the environmental dependence of galaxy
gas-phase metallicity. In \citet{b42} the gas-phase metallicity --
mass relation turns out to be weakly dependent on environment, defined
as the number density of galaxies within the projected distance to the
closest 4th and 5th neighbours. Star-forming galaxies show a marginal
increase of their O/H abundance at fixed stellar mass in denser
environments, equivalent to a few percent for the more massive galaxies
up to $\sim$20$\%$ at log($M_{\star}$/M$_{\odot}) <$ 9.5. The authors
concluded that the evolution of star forming galaxies is mainly driven
by their intrisic properties, and does not depend on environment at
large.  \citet{b43} found a stronger dependence of gas-phase
metallicity on galaxy environment (defined as the number density of
galaxies within the projected distance to the closest 3rd neighbour),
in that galaxies residing in higher density environments are
metal-richer (see also Ellison et al. 2009).  The question arises
whether these differences between \citet{b42} and \citet{b43}
may be due to different definitions of environment.  As already
discussed by e.g. \citet{b44} and recently \citet{b45}, the projected
number density of galaxies is of `tricky' interpretation as it depends
on environment itself. In massive haloes such as clusters, it probes 
environment on a scale smaller than the halo virial radius, while in
low mass haloes the environment probed typically encompasses multiple 
dark matter haloes.

In this paper, we revisit the issue of gas-phase metallicity
vs. galaxy environment by making use of the catalogue of galaxy groups
extracted from the SDSS Data Release 4 by \citet{b46}. In this
catalogue, environment is defined as the total
amount of dark matter associated with each galaxy group, and also
allows us to distinguish (and treat separately) central from
satellite galaxies, for which semi-analytic models of galaxy
formation and evolution predict rather different evolutionary paths.
We aim at achieving a detailed assessment of how
gas-phase metallicities of galaxies depend on galaxy stellar mass
(over the interval 8 $<$ log($M_{\star}/h^{-2}$M$_{\odot}) <$ 11),
hierarchy (centrals vs satellites) and environment across a large range
of halo masses (11 $<$ log($M_{\rm h}/h^{-1}$M$_{\odot}) <$ 14), in
order to gain important insights into how environment regulates
gas inflows feeding the star formation activity of galaxies, and 
affects the efficiency of galactic outflows through the pressure 
applied by the intragroup hot gas.

This paper is organized as follows. In Sect. 2 we describe the
catalogue of galaxy groups obtained by \citet{b46} for the SDSS Data
Release 4 (DR4). The stellar mass - gas-phase metallicity relation
obtained for satellites and centrals separately is discussed in
Sect. 3, while the halo mass - gas-phase metallicity relation is
presented in Sect. 4. The results are discussed in Sect. 5 and the
conclusions follow in Sect. 6.  Throughout this paper we adopt a flat
$\Lambda$CDM cosmology with $\Omega_{\rm m}$ = 0.238 and
$\Omega_{\Lambda}$ = 0.762 (Spergel et al. 2007) and we express units
that depend on the Hubble constant in terms of $h = H_o/$(100 km
s$^{-1}$ Mpc$^{-1}$).

\begin{figure*}
\includegraphics[width=100mm]{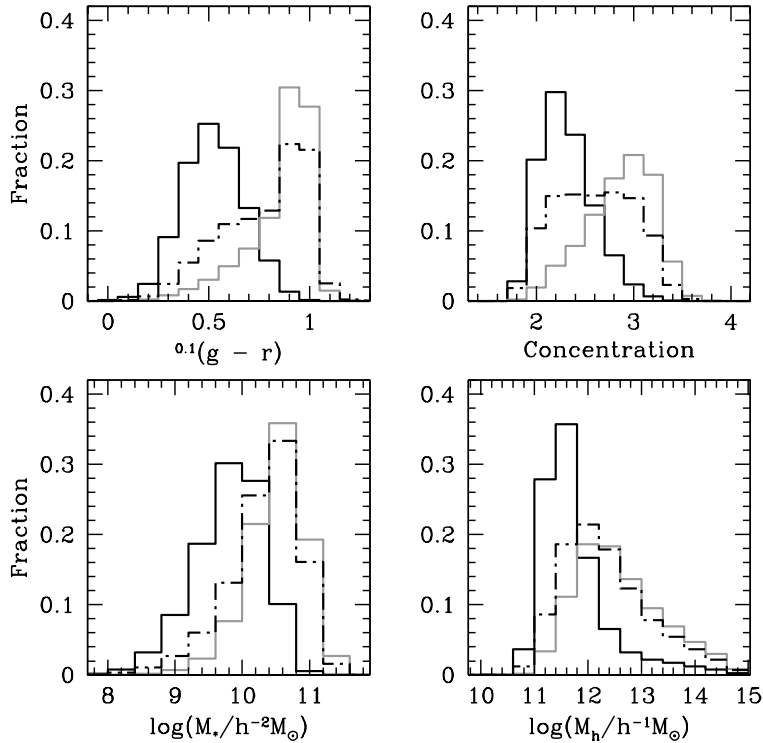}
\caption{The normalized distributions of sample II (black dot-dashed
  line) and samples $S$ and $G$ (solid grey and black, respectively)
  in colour, concentration, stellar mass and halo mass.  The
  distributions are not weighted by 1$/V_{\rm max}$.}
\end{figure*}

\section{DATA}

Our analysis is based on the SDSS DR4 catalogue of galaxy groups
constructed by \citet{b46} by applying the halo-based group finder
algorithm of \citet{b50} to the New York University Value-Added Galaxy
Catalogue (NYU-VAGC, Blanton et al. 2005) extracted from SDSS DR4
(Adelman-McCarthy et al. 2006). From the NYU-VAGC Main Galaxy Sample
\citet{b46} selected all galaxies with an apparent magnitude
(corrected for Galactic foreground extinction) brighter than $r$ = 18
mag, in the redshift range 0.01 $\leq z \leq$ 0.20 and with a redshift
completeness $C_z >$ 0.7. These galaxies were used to build three
galaxy groups samples: sample I, which only uses the 362356 galaxies
with measured redshifts from the SDSS; sample II, which includes an
additional 7091 galaxies with SDSS photometry but redshifts taken from
alternative surveys; sample III which lists an additional 38672 galaxies
lacking a redshift due to fibre collisions, but being assigned the
redshift of their nearest neighbour (cf. Zehavi et al. 2002). This
paper focuses solely on sample II, where galaxies are split into
`centrals' (defined as the most massive group members on the basis of
their $M_{\star}$) and `satellites' (all group members that are not
centrals).  Galaxy magnitudes and colours are based on the standard
SDSS Petrosian technique (Petrosian 1976; Strauss et al. 2002), are
corrected for Galactic foreground extinction (Schlegel, Finkbeiner \&
Davis 1998) and $K$-corrected and evolution corrected to $z$ = 0.1
according to the procedure of \citet{b57}. We use the notation
$^{0.1}M_X$ to indicate the resulting absolute magnitude in the
photometric $X$ band. Galaxy stellar masses are computed using the
relation between stellar mass-to-light ratio and colour of
\citet{b58}, and galaxy morphology is defined in terms of the
concentration parameter ($C$ = $r_{90}/r_{50}$ where $r_{90}$ and
$r_{50}$ are the radii encompassing 90$\%$ and 50$\%$ of the Petrosian
$r-$band flux, respectively).  Values of $C$ larger than 2.6 typically
refer to bulge-dominated galaxies (cf. Strateva et al. 2001).

\begin{figure*}
\includegraphics[width=160mm]{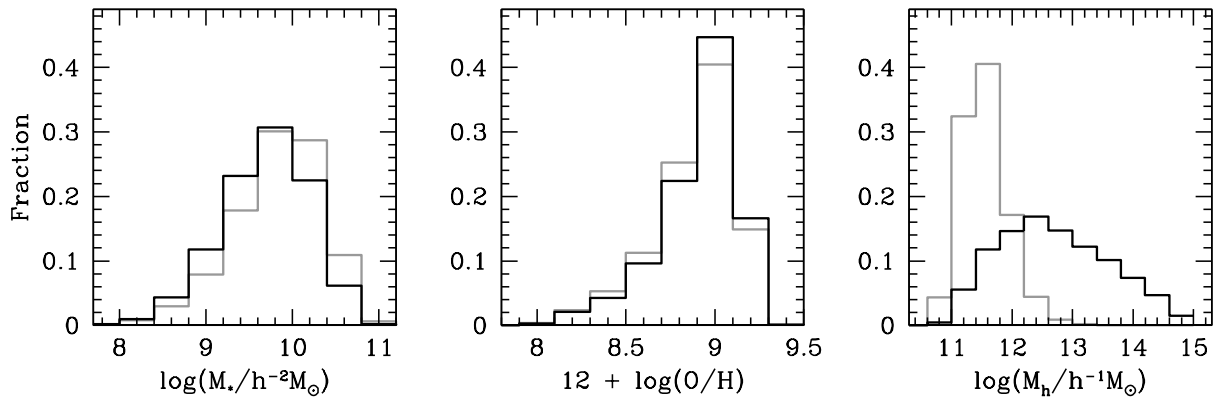}
\caption{The normalized distributions of centrals (grey) and
  satellites (black) in sample $G$ in stellar mass, gas-phase
  metallicity and halo mass.  The distributions are not weighted by
  1$/V_{\rm max}$. }
\end{figure*}

For each group in the catalogue \citet{b46} provided two estimates of
its dark matter halo mass, $M_{\rm h}$, one based on the ranking of
its total luminosity, the other on the ranking of its total stellar
mass. These halo masses deviate from each other by about 0.1 dex at
low $M_{\rm h}$ and 0.05 dex at the massive end. The method of
\citet{b46} is able to assign $M_{\rm h}$ only to groups more massive
than $\sim 10^{12} h^{-1}$ M$_{\odot}$ and with one or more members
brighter than $^{0.1}M_r -$ 5log$h =$ -19.5 mag. For all the other
groups, \citet{b60} used the relations between the luminosity (or
$M_{\star}$) of central galaxies and the halo mass of their groups to
estimate the halo mass of single central galaxies down to $M_{\rm h}
\sim$ 10$^{11} h^{-1}$ M$_{\odot}$. In what follows we will use the
halo masses derived from the group's total stellar mass, because, as
shown by \citet{b61}, stellar mass is a better indicator of halo mass
than luminosity. Since the groups catalogue of \citet{b46} is not
volume limited, it suffers from the Malmquist bias, which causes an
artificial increase of the average luminosity (and $M_{\star}$) of
galaxies with increasing redshift. This effect is especially important
for satellites, since they span a large range of stellar masses within
each halo. We take the Malmquist bias into account by weighting each
galaxy by 1$/V_{\rm max}$, where $V_{\rm max}$ is the comoving volume
of the Universe out to a comoving distance at which a galaxy would
still have made the selection criteria of the groups
sample. Therefore, all distributions in the following sections are
weighted by 1$/V_{\rm max}$ unless specified otherwise.

\subsection{Stellar population parameters: sample $S$}

We matched sample II with the catalogue of stellar ages and
metallicities obtained by \citet{b62} for the galaxies in the SDSS
DR4. Briefly, \citet{b62} compared the strength of stellar absorptions
in the observed spectrum of each galaxy with the predictions of a
Monte Carlo library of 150 000 star formation histories (SFHs), based
on the \citet{b63} population synthesis code and the \citet{b64}
initial mass function. The SFHs in the library have a declining star
formation rate (SFR) over varying timescales, and are superposed with
random bursts of varying intensity and duration. These bursts are
added in such a way that only 10$\%$ of the models can experience a
burst of star formation in the last 2 Gyr. For each galaxy a
probability density distribution of stellar $r-$band flux-weighted
stellar ages and metallicities can be derived, whose median values we
will use for our analysis below. It has to be noted that the stellar
ages and metallicities derived as above refer to the redshift at which
galaxies are observed. No correction to $z$ = 0 was attempted because
this would require an accurate knowledge of the SFH from the redshift
of the observations to the present. The uncertainty on the stellar
ages and metallicities depends on the spectral signal-to-noise (S/N);
a S/N $\geq$ 20 ensures $\Delta$log($Z$) $<$ 0.3 dex,
$\Delta$log(Age) $<$ 0.2 dex and an average uncertainty on both
parameters of $\sim$0.12 dex.

All galaxies in sample II which have a spectral S/N $\geq$ 20 and a
stellar age and metallicity estimates computed as in  \citet{b62} hereafter
constitute our sample $S$. This sample was extensively studied by \citet{b65},
who found that satellites are older and metal richer than centrals of
the same $M_{\star}$, and this difference increases with decreasing
$M_{\star}$. In addition, the average age and stellar metallicity of
low-mass satellites ($M_{\star} \leq$ 10$^{10} h^{-2}$ M$_{\odot}$)
increase with the mass of the halo in which they reside. The authors
interpreted these trends as due to the quenching of star formation in
satellites, which leaves their stellar populations to evolve passively
(thus explaining their older ages), and to tidal stripping, which
removes a non-negligible fraction of their stellar mass. In this picture,
the stellar metallicity of present-day satellites reflects the 
maximum stellar mass that they reached during their lifetime and which
can be significantly higher than their present-day $M_{\star}$.

\subsection{Gas metallicities and star formation rates: sample $G$}

Sample II was also cross-matched with the catalogue of gas-phase
metallicities [i.e. 12 $+$ log(O/H)] published by \citet{b7} for the
galaxies in SDSS DR4. The analysis performed by \citet{b7} can briefly
be summarized in the following steps: {\it i)} the continuum emission
in each galaxy spectrum was modelled with a combination of single
stellar populations from \citet{b63} of the same stellar metallicity,
also convolved with the extinction law of \citet{b68}, and
subsequently subtracted from the observed spectrum. {\it ii)} The
lines in the resulting pure emission-lines spectrum were
simultaneously fitted with Gaussians by imposing that all the Balmer
lines have the same width and velocity offset, and likewise for the
forbidden lines. {\it iii)} The gas-phase metallicity was
statistically estimated for all star forming galaxies by fitting the
more prominent emission lines with the models of \citet{b66}.  These
models combine the synthetic stellar populations of \citet{b63} with
the photoionization code CLOUDY (Ferland 1996) and the \citet{b68}
extinction law, so that the galaxy ISM conditions (e.g. gas
ionization, metallicity and dust attenuation) are intertwined with the
radiation field and nuclear synthesis of the underlaying stellar
populations. For each galaxy \citet{b7} derived a likelihood
distribution of the gas-phase metallicity from which we extracted
the median, the 16th and 84th percentile values. All galaxies in
sample II whose H$\beta$ equivalent width is EW (H$\beta \leq$ -3
\AA\/ (corresponding to a S/N $\geq$ 10) and whose gas-phase
metallicity is available from \citet{b7} make up our sample $G$.

For the galaxies in sample $G$ we retrieved the median values of
stellar mass, $m_{\rm fib}$, star formation rate, SFR$_{\rm fib}$, and
specific SFR$_{\rm fib}$ (sSFR$_{\rm fib}$ = SFR$_{\rm fib}$/$m_{\rm
  fib}$) in the fibre, the median values of global star formation
rate, SFR$_{\rm glo}$, and specific SFR$_{\rm glo}$ (sSFR$_{\rm glo}$
= SFR$_{\rm glo}$/$M_{\star}$) from the SDSS Data Release 7. For star
forming galaxies (like those in sample $G$), the SFR$_{\rm fib}$ was
directly computed from their emission lines following the procedure of
\citet{b8}, who modelled the observed emission lines with the models
of \citet{b66}. The global star formation rate of the galaxies in DR7
was derived by fitting the photometry of the outer regions of galaxies
with the synthetic stellar population models of \citet{b63}
convolved with the extinction law of \citet{b68} (cf. Salim et
al. 2007).  Using this same procedure, \citet{b8} derived the global
and fibre stellar masses, $M_{\star}$ and $m_{fib}$, from the galaxy
integrated magnitudes and the magnitudes measured within the fibre.

\begin{figure*}
\includegraphics[width=100mm]{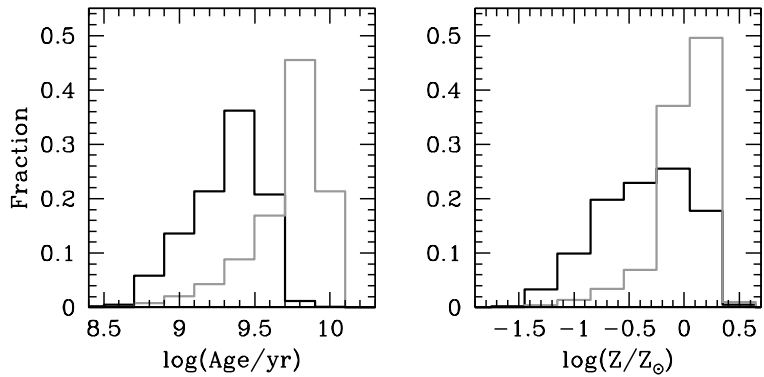}
\caption{The normalized distributions of galaxies in sample $S$ (grey)
  and sample $C$ (black) in stellar age and metallicity. The
  distributions are not weighted by 1$/V_{\rm max}$.}
\end{figure*}

\subsection{Combined sample}

We extracted from sample $G$ all galaxies for which the spectral S/N
is larger than 20, and the stellar age and metallicity are available
together with their gas-phase metallicity and star formation rate.  We
refer to this subset of objects as the composite sample, i.e. sample
$C$. Table 1 summarizes the number of central and satellite galaxies
in the samples $S$, $G$ and $C$ defined above. In Fig. 1 we compare
the normalized distributions of the galaxies in sample II (black
dashed line), sample $S$ (grey solid line) and $G$ (black solid line)
in colour, concentration, stellar and halo masses. We note that the
sample definition introduces a clear separation in colour and
concentration between sample $G$ and sample $S$. Being star forming,
the galaxies in sample $G$ turn out to be bluer and more disk-like
(with lower concentrations) than those in sample $S$. As already shown
by \citet{b47}, star forming galaxies have on average lower
$M_{\star}$ and preferentially reside in low-mass haloes.  This trend
is mainly due to the central galaxies of sample $G$ as it can be
inferred from the right-hand panel of Fig. 2, while the satellites in the
same sample span a wide range of environments from $M_{\rm h} \simeq$
10$^{11}$ to 10$^{15}$ $h^{-1}$M$_{\odot}$. In addition, centrals seem
to be slightly more massive and metal-poorer than satellites in sample
$G$ (left-hand and middle panels of Fig. 2). It is also interesting to
compare sample $S$ and sample $C$ in stellar age and metallicity; we
find in Fig. 3 that the star forming galaxies in sample $C$ have on
average younger and metal-poorer stellar populations than those in
sample $S$. We are thus probing different ranges of galaxy properties
with respect to the analysis in Pasquali et al. (2010).

\section{The stellar mass - gas-phase metallicity relation}

The dependence of gas-phase metallicity on stellar mass is plotted in
Fig. 4 for centrals (grey) and satellites (black). We computed the
median (solid lines), the 16th and 84th percentile values (dashed
lines) of the gas-phase metallicity distribution of centrals and
satellites in sample $G$ per bin of $M_{\star}$. It turns out that
centrals and satellites follow a qualitatively similar relation
between gas-phase metallicity and $M_{\star}$, whereby the metallicity
of their gas increases with their stellar mass.  The gas in satellites
is, though, metal richer than in centrals at nearly all stellar
masses, and this is true for the median values as well as for the 16th
and 84th percentiles.

The difference in the median 12 $+$ log(O/H) between satellites and
centrals decreseas from 0.06 dex at
log$(M_{\star}/$h$^{-2}$M$_{\odot}) \simeq$ 8.25 to 0.004 dex at
log$(M_{\star}/$h$^{-2}$M$_{\odot}) \simeq$ 10.75. This trend holds
also for volume limited sub-samples extracted from sample $G$.

For each bin of stellar mass, we constructed the cumulative
distributions in 12 $+$ log(O/H) of centrals and satellites, and used
the Kolmogorov - Smirnov (KS), two sample test to assess the
statistical significance of the metallicity offset seen in Fig. 4. The
probability at which the null hypothesis (i.e. centrals and satellites
are drawn from the same population) is rejected is larger than 95$\%$
in the range 8 $\leq$ log($M_{\star}/$h$^{-2}$M$_{\odot}) \leq$ 10.5,
and drops to $\sim$19$\%$ for galaxies more massive than $M_{\star}$ =
10$^{10.5}$ h$^{-2}$M$_{\odot}$. Thus, over the stellar mass range
spanned by our sample, the cumulative distributions indicate that
satellites have a higher gas-phase metallicity than centrals.

\begin{figure}
\includegraphics[width=70mm]{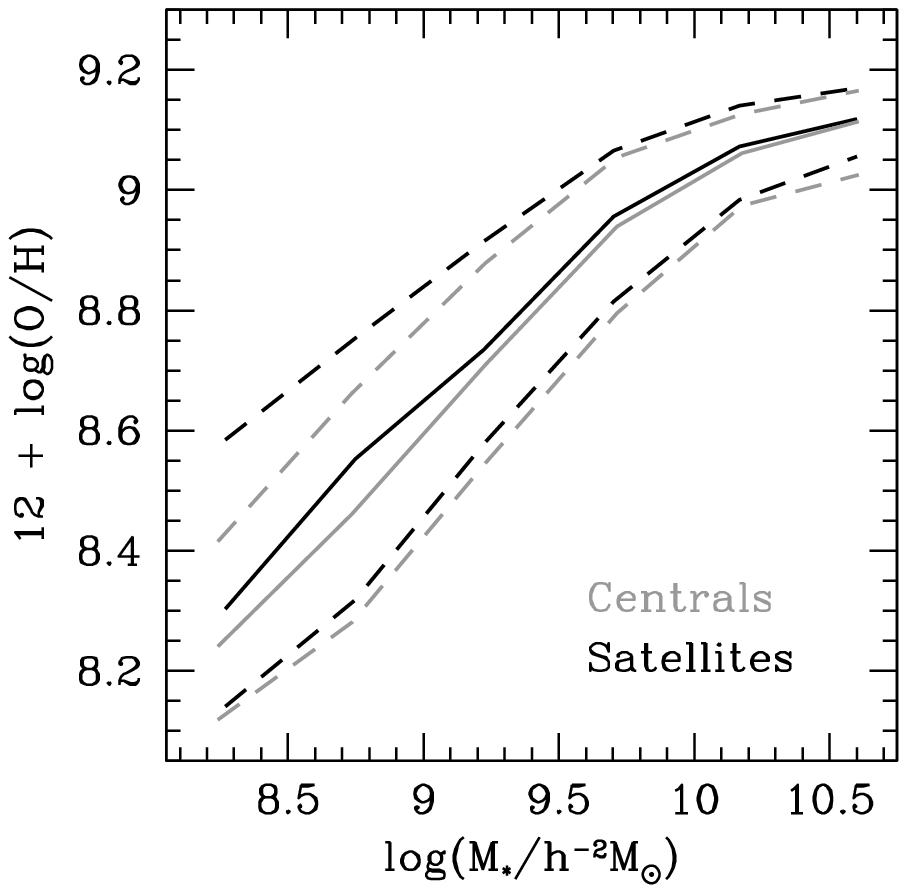}
\caption{The $M_{\star}$ - gas-phase metallicity relation for central
  (grey) and satellite (black) galaxies. The solid lines represent the
  median metallicity, while the dashed lines the 16th and 84th
  percentiles of the metallicity distribution of centrals and
  satellites in each bin of stellar mass. Centrals and satellites
  belong to sample $G$.}
\end{figure}

In order to check whether/how the $M_{\star}$ - gas-phase metallicity
relation of satellites depends on environment, we derived it for
satellites split among different bins of halo mass.  The results are
shown in Fig. 5, where satellites residing in different groups are
colour coded; the black dashed line traces the median 12 $+$ log(O/H)
of centrals from Fig. 4 and the grey band represents the 16th - 84th
percentile range of the distribution of centrals also from Fig. 4.
The 1$\sigma$ errorbars represent the error on the mean gas-phase
metallicity weighted by 1/$V_{\rm max}$, and were computed by
propagating the uncertainties associated with the individual
measurements of 12 $+$ log(O/H).  Satellites less massive than
log($M_{*}/h^{-2}$M$_{\odot}) \simeq$ 9.3 and in groups more massive
than log($M_{\rm h}/h^{-1}$M$_{\odot}) \simeq$ 12 appear to follow the
same $M_{\star}$ - gas-phase metallicity relation.  The overall shift
to higher gas-phase metallicities of satellites with respect to
centrals is due only to satellites in haloes more massive than
10$^{12}h^{-1}$M$_{\odot}$, with their stellar mass - gas-phase
metallicity relation slightly steepening as $M_{\rm h}$ increases.

Satellites in the least massive groups (11 $<$ log($M_{\rm
  h}/h^{-1}$M$_{\odot}) <$ 12) and more massive than
log($M_{\star}/h^{-2}$M$_{\odot})$ = 9.3 are very similar to centrals
in terms of their gas-phase metallicity, while less massive satellites
exhibit higher 12 $+$ log(O/H) values than centrals as their stellar
mass decreases [the maximum difference, $+$0.11 dex, is registered at
log($M_{\star}/h^{-2}$M$_{\odot}) \simeq$ 8.2].

\begin{figure}
\includegraphics[width=70mm]{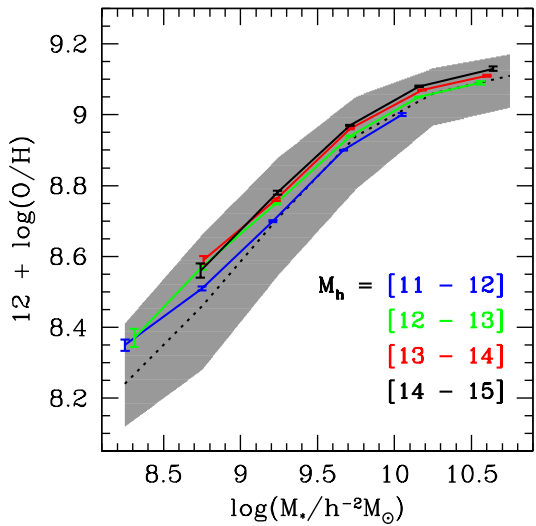}
\caption{The $M_{\star}$ - gas-phase metallicity relation for centrals
  (grey) and satellites (colour) split among different bins of halo mass. The
  black dashed line represents the median 12 $+$ log(O/H) of centrals,
  while the grey band marks the 16th - 84th percentile range of the
  distribution of centrals.}
\end{figure}

\section{The halo mass - gas-phase metallicity relation}

We show in Fig. 6 the dependence of the gas-phase metallicity of
central and satellite galaxies on the halo mass of the group in which
they reside. The grey and black lines refer to centrals and
satellites, respectively; the solid and dashed lines represent the
50th and 16th/84th percentiles of the gas-phase metallicity
distribution within each bin of M$_{\rm h}$. In halos with
  $M_{\rm h} \lta 10^{12} h^{-1} M_{\odot}$ the gas-phase metallicity
  of central galaxies increases sharply with halo mass.  Centrals in
  more massive halos, though, have roughly constant gas-phase
  metallicities. Note that the small number statistics for massive
  groups in the Yang et al. (2007) catalogue prevent us from probing
  this halo mass dependence beyond $M_{\rm h} \sim 10^{14}
  h^{-1}M_{\odot}$. Regardless of halo mass, central galaxies are
metal richer than satellites in halos of the same mass by
$\sim$0.5 dex on average. This is simply a reflection of the fact
  that, by definition, central galaxies are the more massive galaxies
  in their groups, combined with the fact that gas-phase metallicity
  is an increasing function of stellar mass. The oxygen abundance of
satellites smoothly increases with halo mass up to $M_{\rm h} \simeq$
10$^{13} h^{-1}$M$_{\odot}$, and is constant in more massive groups.
Overall, it is seen increasing by $\sim$0.5 dex across the full range
of environments probed by sample $G$, in agreement with the findings
of \citet{b43}.

We split the satellites in sample $G$ among different bins of stellar
mass and computed their $M_{\rm h}$ - gas-phase metallicity relation
as a function of $M_{\star}$. The results are illustrated in the left
hand panel of Fig. 7, where satellites are colour coded on the basis
of their stellar mass. The black dashed line traces the median 12 $+$
log(O/H) of centrals as a function of $M_{\rm h}$ (from Fig. 6), while
the grey band indicates their 16th-to-84th percentile distribution in
gas-phase metallicity as shown in Fig. 6.  The 1$\sigma$ errorbars
for satellites
represent the error on the mean gas-phase metallicity weighted by
1/$V_{\rm max}$, and were computed by propagating the uncertainties
associated with the individual measurements of 12 $+$ log(O/H).  The
offset in 12 $+$ log(O/H) among the different $M_{\rm h}$ - gas-phase
metallicity relations is simply due to stellar mass.  We see that the
gas-phase metallicity of satellites increases with their halo mass in
all bins of $M_{\star}$. The amplitude of this trend across the full
$M_{\rm h}$ range amounts to $\sim$0.15 dex for satellites with
log($M_{\star}/h^{-2}$M$_{\odot}) <$ 9, $\sim$0.1 dex for those with 9
$<$ log($M_{\star}/h^{-2}$M$_{\odot}) <$ 10.5 and only $\sim$0.02 dex
for the most massive satellites. These findings are consistent with
the results of \citet{b42}.  To better illustrate the dependence of
gas-phase metallicity on $M_{\rm h}$, we computed the slope d[12 $+$
log(O/H)]$/$d[log($M_{\rm h}$)] in each stellar mass bin, and plotted
it as a function of $M_{\star}$ in the right hand panel of Fig. 7.  A
weak trend can be recognized whereby the strength of the $M_{\rm h}$ -
gas-phase metallicity relation decreases with increasing stellar mass.

\begin{figure}
\includegraphics[width=70mm]{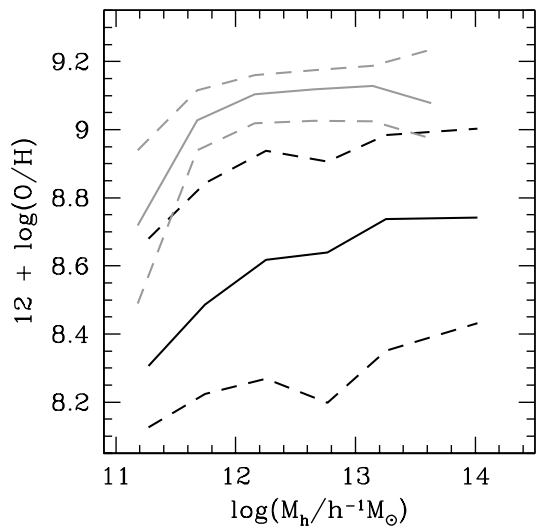}
\caption{The $M_{\rm h}$ - gas-phase metallicity relation for central
  (grey) and satellite (black) galaxies. The solid lines represent the
  median metallicity, while the dashed lines the 16th and 84th
  percentiles of the metallicity distribution of centrals and
  satellites in each bin of stellar mass. Centrals and satellites
  belong to sample $G$.}
\end{figure}

\begin{figure*}
\includegraphics[width=150mm]{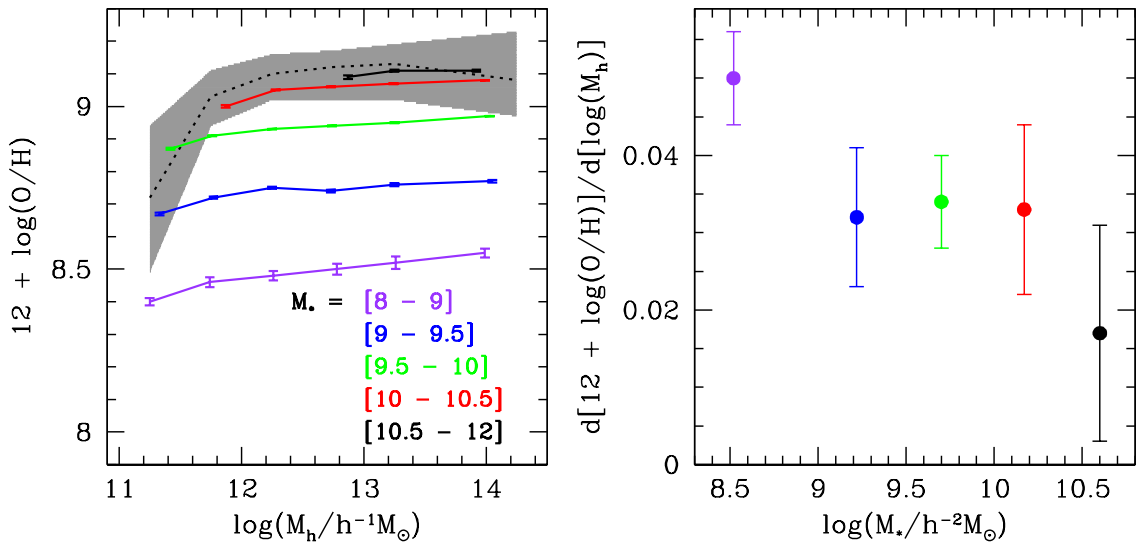}
\caption{{\it Left:} the $M_{\rm h}$ - gas-phase metallicity relation
  for satellite galaxies (colour) split among different bins of stellar
  mass. The black dashed line represents the median metallicity of
  centrals, while the grey band shows the 16th - 84th percentiles
  distribution of centrals as a function of halo mass. Centrals and
  satellites belong to sample $G$. {\it Right:} the slope of the
  $M_{\rm h}$ - gas-phase metallicity relations shown in the left hand
  panel is plotted as a function of stellar mass. The errorbars
  indicate the 1$\sigma$ uncertainty.}
\end{figure*}

\section{Discussion}

In the previous sections we have shown that the gas-phase metallicity
of central and satellite galaxies depends primarily on their stellar
mass. In agreement with the results of \citet{b7}, their gas becomes
metal richer as their stellar mass increases. However, at fixed
$M_{\star}$ the gas-phase metallicity is higher in satellites than in
centrals, and this observed offset increases with 
  decreasing stellar mass. In addition, at fixed $M_{\star}$ the
oxygen abundance of satellites is found to be higher in more massive
haloes. This trend is observed for satellites less massive than
$M_{\star} \simeq$ 10$^{10.5}h^{-2}$M$_{\odot}$ with an amplitude
  of $\sim 0.1$dex over the range 11 $<$ log($M_{\rm
  h}/h^{-1}$M$_{\odot}) <$ 14, and becomes considerably weaker for the
most massive satellites. Hence, there is a clear and `pure'
  dependence of the gas-phase metallicity of satellites on
  environment, but this dependence is much weaker than the dependence
  on stellar mass. In what follows we investigate the physical
mechanisms that may be responsible for the ($M_{\star}$,$M_{\rm h}$) -
gas-phase metallicity relations of satellite galaxies.
  
\subsection{Gas-phase and stellar metallicities}

We start by comparing the properties in common between the galaxies in
sample $C$ and those in sample $S$.  We recall that sample $C$ is
built from sample $S$ and includes only galaxies with gas-phase
metallicity, stellar age and metallicity, while sample $S$ contains
galaxies with measured stellar age and metallicity, regardless of
whether their gas-phase metallicity is available (see
Sect. 2). Similarly to sample $G$ (cf. Fig. 4), also the satellites in
sample $C$ display a higher median oxygen abundance when
compared with equally massive centrals. Figure 8 shows that their
difference in median 12 $+$ log(O/H) at fixed $M_{\star}$ decreases
from $\sim$0.14 dex at $M_{\star} <$ 10$^{9}h^{-2}$M$_{\odot}$ to
$\sim$0.02 dex at $M_{\star} <$ 10$^{10.3}h^{-2}$M$_{\odot}$.
  Note that this difference is larger than that measured between
  centrals and satellites in sample $G$ as shown in Fig. 4. This is likely to be a
  consequence of the fact that sample $C$ is biased towards galaxies
with weaker emission lines and lower star formation activity than
those in sample $G$.  In this respect, satellites in sample $C$ may
have been more significantly quenched by their environment than
satellites in sample $G$. We caution, though, that sample $C$
  suffers from small number statistics when compared to sample $G$.
The KS two sample test performed on the cumulative distributions in 12
$+$ log(O/H) of centrals and satellites in sample $C$ in different
bins of stellar mass establishes that satellites are metal richer than
centrals at a confidence level between 78$\%$ and 100$\%$ in the range
8 $<$ log($M_{\star}/h^{-2}$M$_{\odot}) <$ 10.5, while the most
massive satellites can not be distinguished from equally massive
centrals.

\begin{figure}
\includegraphics[width=70mm]{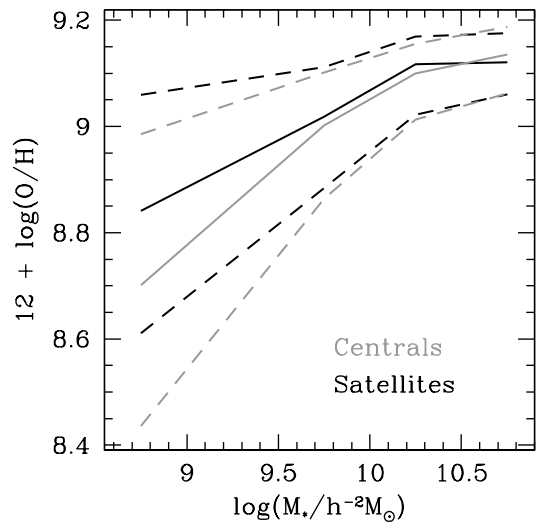}
\caption{The $M_{\star}$ - gas-phase metallicity relation for central
  (grey) and satellite (black) galaxies in sample $C$, i.e. with
  measures of both gas and stellar metallicity. The solid lines
  represent the median metallicity, while the dashed lines the 16th
  and 84th percentiles of the metallicity distribution of centrals and
  satellites in each bin of stellar mass.}
\end{figure}

Using sample $S$, \citet{b65} found that satellites have older,
luminosity-weighted ages and higher stellar metallicities than equally
massive centrals. Both these differences become smaller for more
  massive galaxies. In addition, they showed that satellites with
  $M_{\star} \lta 10^{10}h^{-2}M_{\odot}$ have ages and metallicities
  that increase with the mass of the host halo in which they reside.
\citet{b65} explained these trends by invoking {\it i)} strangulation,
i.e. the removal of the gas reservoir of satellites after they are
accreted by their host halo. The consequent quenching of star
formation leaves the stellar populations of satellites to evolve
passively, while the more prolonged star formation activity of
centrals keeps their luminosity-weighted age younger. {\it ii)} Tidal
stripping, which causes satellites to lose stellar mass once they have
been accreted onto their host group. The fact that present day
satellites descend from more massive progenitors is left imprinted in
their stellar metallicity, which is not expected to change
significantly after a satellite is accreted onto a bigger halo and its
star formation activity is quenched. The effects of strangulation and
tidal stripping likely depend on the time of infall, i.e. when a
galaxy became a satellite for the first time.  The earlier the infall
time is, the older and less massive a satellite is, and the more
massive the group in which the satellite resides (given that
more massive haloes assemble earlier).  This dependence on
  infall time naturally explains why the ages and stellar
  metallicities of satellites increase with $M_{\rm h}$.

\begin{figure*}
\includegraphics[width=120mm]{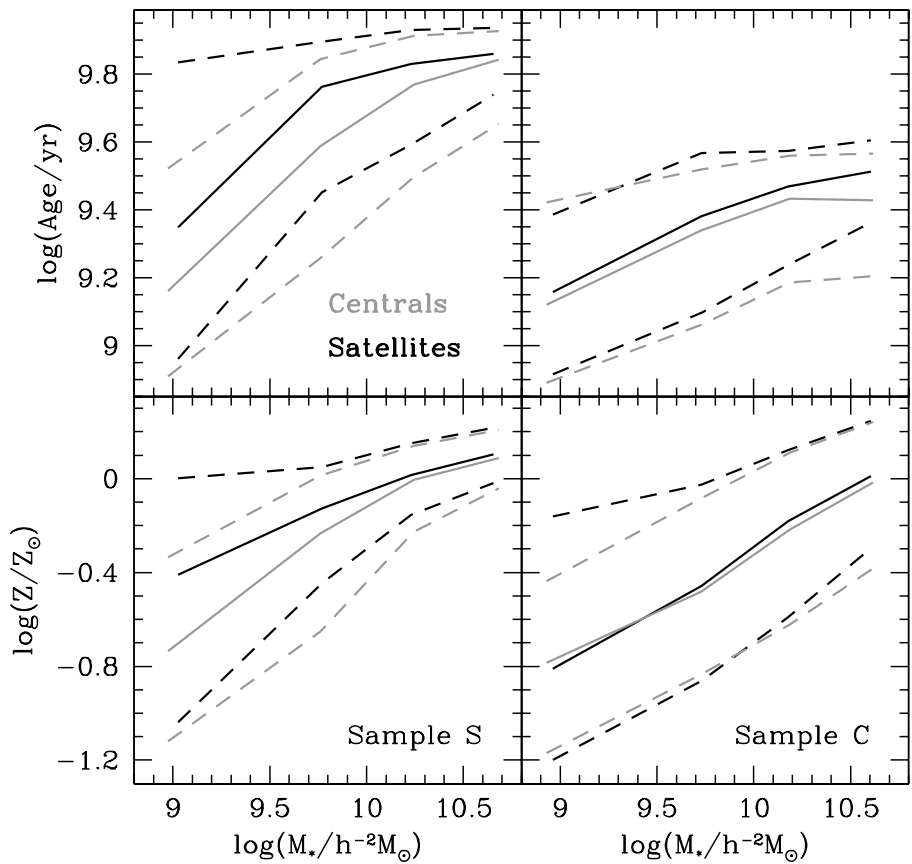}
\caption{The distributions of stellar age and metallicity as a
  function of stellar mass for centrals (grey) and satellites (black)
  beloging to sample $S$ (left hand panels) and to sample $C$ (right
  hand panels). The solid lines represent the median age/metallicity,
  while the dashed lines the 16th and 84th percentiles of the
  age/metallicity distribution of centrals and satellites in each bin
  of stellar mass.}
\end{figure*}

Since both stellar and gas-phase metallicities correlate with
$M_{\star}$, one might also expect the higher gas-phase
  metallicities of satellites (with respect to centrals of the same
  $M_{\star}$) to be a manifestation of tidal stripping. In this
  scenario, satellites have higher gas-phase metallicities than
  centrals of the same {\it present-day} mass, simply because they had
  a higher stellar mass at infall: the gas-phase metallicity serves as
  an indicator of the maximum stellar mass the galaxy reached during
  its lifetime.  In order to test this hypothesis, we compare in
Fig. 9 the $M_{\star}$ - stellar age and the $M_{\star}$ - stellar
metallicity relations of centrals and satellites in sample $S$ with
those of centrals and satellites in sample $C$. As already pointed out
by Fig. 3, the galaxies in sample $C$ are characterized by younger
stellar ages and lower stellar metallicities than those in sample $S$.
At fixed stellar mass satellites are systematically older than
centrals in both samples. This age difference is rather small for
sample $C$ (about 0.04 dex at M$_{\star} <$
10$^{10}h^{-2}$M$_{\odot}$), and statistically significant only for
galaxies more massive than 10$^{9.5}h^{-2}$M$_{\odot}$. In fact, the
KS two sample test applied to the age cumulative distributions of
sample $C$ indicates that centrals and satellites with
log(M$_{\star}/h^{-2}$M$_{\odot}) >$ 9.5 are drawn from two different
populations at a $>$92$\%$ confidence level, while less massive
centrals and satellites belong to the same parent population. In
sample $S$ the age difference between satellites and centrals is
larger, and decreases from 0.19 dex at $M_{\star} <$
10$^{9}h^{-2}$M$_{\odot}$ to 0.02 dex at $M_{\star} <$
10$^{10.5}h^{-2}$M$_{\odot}$. The KS two sample test applied to the
age cumulative distributions of centrals and satellites in different
$M_{\star}$ bins shows that the hypothesis of centrals and satellites
belonging to the same population is rejected at a $>$99$\%$
confidence level. We may thus conclude that the satellites in sample
$C$ were accreted later and their star formation activity has been
quenched less than the satellites in sample $S$. Consequently, they
should have also experienced less tidal stripping, so that their
stellar metallicities should be closer to those of equally massive
centrals. This is indeed the case; the bottom panels of Fig. 9
  show that satellites in sample $C$ have stellar metallicities that
  are basically indistinguishable from those of centrals of the same
  stellar mass. The differences are less than 0.02 dex, implying that
  satellites in sample $C$ have lost $<$10$\%$ of their pre-infall
  stellar mass. For comparison, the difference in the median
log($Z/Z_{\odot}$) between satellites and centrals in sample $S$
decreases from $\sim$0.3 dex at M$_{\star} <$
10$^{9}h^{-2}$M$_{\odot}$ to $<$0.02 dex at M$_{\star} >$
10$^{10}h^{-2}$M$_{\odot}$, indicating that satellites in
  sample $S$ have lost between $\sim$75$\%$ (for a present day
$M_{\star}$ = 10$^{9}h^{-2}$M$_{\odot}$) and $\sim$20$\%$ (for a
present day $M_{\star}$ = 10$^{10.5}h^{-2}$M$_{\odot}$) of their
stellar mass at infall. To summarize, Fig. 9 indicates that satellites
in sample $C$ have not suffered significant tidal stripping (yet), but
have undergone quenching of their star formation activity at some,
small degree. Therefore, we conclude that stellar mass stripping
  cannot explain why satellites have higher gas-phase metallicities
  than centrals of the same stellar mass.
  
In further support of this conclusion, we directly compare the gas-phase 
metallicities of centrals and satellites at fixed stellar metallicity in Fig. 10. 
If the same stellar stripping responsible for creating the (small) offset in stellar 
metallicity between star-forming centrals and satellites were the only responsible 
for the observed offset in gas-phase metallicities, we would expect the relation 
between gas-phase and stellar metallicity to be the same for central and satellite 
galaxies. Figure 10 shows that clearly this is not the case: at log($Z/Z_{\odot}) <$ -0.4 (corresponding to $M_{\star} <$ 10$^{9.5}h^{-2}$M$_{\odot}$, see Fig. 9) 
satellites have higher gas-phase metallicities than centrals of the same stellar 
metallicity. Rather, Fig. 10 suggests that the 
differences in gas-phase metallicity between centrals and satellites have their origin 
in something that is specific to the gas and that has a larger impact on less chemically 
evolved (hence preferentially low-mass) galaxies. The scatter between gas-phase and
stellar metallicity has been shown to be associated with variations in stellar surface mass 
density, star formation rate and gas mass fraction (Gallazzi et al. 2005).  In the next sections
we test whether any of these parameters may indeed be associated with the difference in 
gas-phase metallicity between centrals and satellites.

\begin{figure}
\includegraphics[width=70mm]{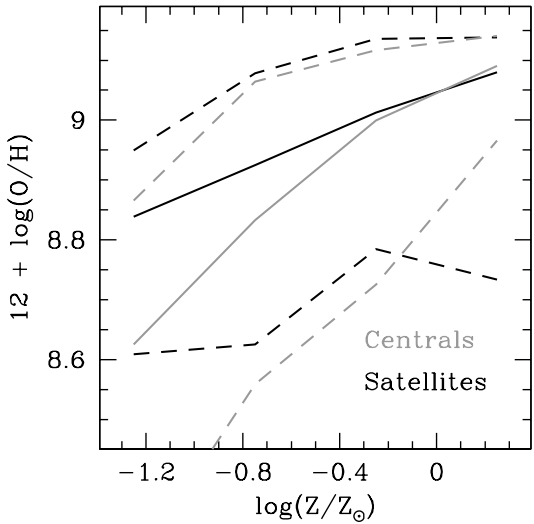}
\caption{The dependence of the gas-phase metallicity of
  centrals (grey) and satellites (black) in sample $C$ on their
  stellar metallicity. The solid lines trace the median 12 $+$
log(O/H), while the dashed lines the 16th and 84th percentiles of
the gas-phase metallicity in each bin of stellar metallicity.}
\end{figure} 

\subsection{Surface mass densities}

\citet{b70} have shown that the star formation activity of low
mass galaxies depends more on their surface mass density than on
their stellar mass. In a naive closed-box model for chemical
  evolution, this would imply that, at fixed $M_{\star}$, galaxies
with a higher surface mass density have transformed a larger amount of
gas into stars and hence increased their gas-phase metallicity. For
the galaxies in sample $G$ we define the surface mass density in the
fibre as $\Sigma_{\rm fib} = m_{\rm fib}/area_{\rm fib}$, where
$area_{\rm fib}$ is the physical area covered by the SDSS fibre (3$''$
in diameter) at the galaxy distance. We use $\Sigma_{\rm fib}$ as a
proxy for the past star formation history that took place over
  the same (central) regions from which the gas-phase metallicities
  have been measured. The solid lines in the left hand panel of
  Fig. 11 plots the median $\Sigma_{\rm fib}$ of centrals (grey) and
  satellites (black) as a function of their stellar mass; the dashed
  lines indicate the corresponding 16th and 84th percentiles.  The
difference in $\Sigma_{\rm fib}$ between satellites and centrals is
$\sim$0.02 dex at most, and changes from being negative at
log($M_{\star}/h^{-2}$M$_{\odot}) <$ 9.5 ($\Sigma_{\rm fib}$ is lower
in satellites) to positive at higher $M_{\star}$.  In the right hand
panel of Fig. 11 the median oxygen abundance (solid line) of centrals
(grey) and satellites (black) is shown as a function of their
surface mass density; as expected from the findings of \citet{b70},
the gas-phase metallicity increases with $\Sigma_{\rm fib}$ for
  both centrals and satellites.  When matched in $\Sigma_{\rm fib}$,
satellites with log($\Sigma_{\rm fib}) <$ 8.5 display a higher
gas-phase metallicity than centrals in the median and 16th and 84th
percentiles values.  We therefore conclude that the offsets in 12
  $+$ log(O/H) are not associated with any offsets in surface mass
  density, as expected in a simple closed box model if the different
  gas-phase metallicities of centrals and satellites owe to
  differences in their star formation histories.

\begin{figure*}
\includegraphics[width=130mm]{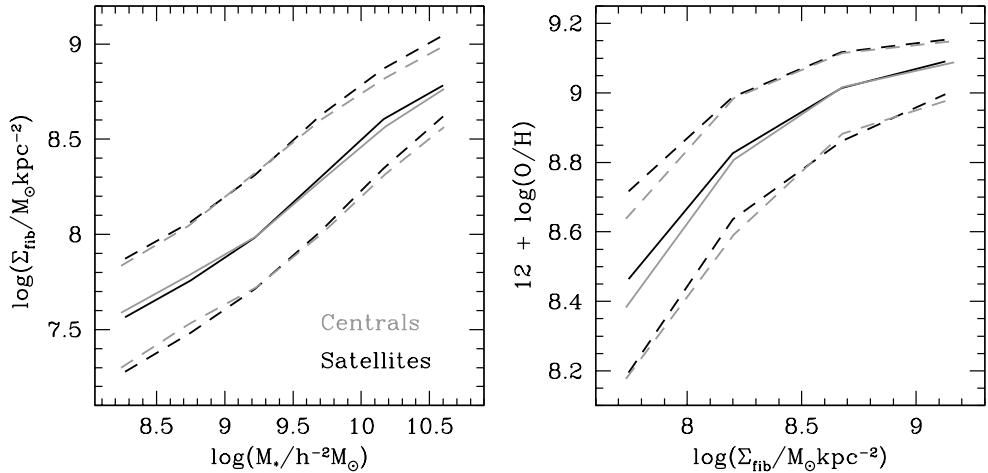}
\caption{{\it Left:} the surface mass density in the fibre as a
  function of stellar mass for centrals (grey) and satellites
  (black). The solid lines show the median $\Sigma_{\rm fib}$ while
  the dashed lines the 16th and 84th percentiles in different bins of
  stellar mass.  {\it Right:} the dependence of gas-phase metallicity
  on $\Sigma_{\rm fib}$ for central (grey) and satellite (black)
  galaxies.}
\end{figure*}

\subsection{Gas mass fractions}

Similarly to $\Sigma_{\rm fib}$, the gas mass fraction can be
considered an indirect measurement of the past star formation activity
undergone by galaxies.  It has been shown by \citet{b77} that the HI
gas mass fraction of galaxies decreases along the gas-phase
metallicity - stellar mass relation as both $M_{\star}$ and the oxygen
abundance increase. Moreover, at fixed stellar mass, gas - poor
galaxies exhibit higher gas-phase metallicities. This poses the
  question whether the difference in gas-phase metallicity between
central and satellite galaxies reflects a difference in their gas mass
fractions. In order to test this, we computed the gas mass
fraction in the fibre for central and satellite galaxies using
Eq. (5) of \citet{b7}; i.e., we derive the surface gas mass
density, $\Sigma_{\rm gas}$, from the star formation surface density
in the fibre by inverting the Schmidt law given by \citet{b73}. The
gas mass fraction in the fibre is then defined as $\mu_{\rm gas} =
\Sigma_{\rm gas}/(\Sigma_{\rm gas} + \Sigma_{\rm fib})$, where
$\Sigma_{\rm fib}$ is the surface mass density calculated in
Sect. 5.2.  The gas mass fraction is plotted as a function of
$M_{\star}$ in the left hand panel of Fig. 12, where centrals are in
grey and satellites in black. Solid lines represent the median
$\mu_{\rm gas}$ in bins of stellar mass, while the dashed lines
  indicate the the 16th and 84th percentiles. The gas mass fraction
decreases as galaxies become progressively more massive, and no
significant difference is seen between centrals and satellites. We
also determined the 16th, 50th and 84th percentiles of the gas-phase
metallicity of centrals and satellites in different bins of $\mu_{\rm
  gas}$, as shown in the right hand panel of Fig. 12. For both
  centrals and satellites, the oxygen abundance decreases with
increasing gas mass fraction, in agreement with the findings of
\citet{b77}, \citet{b75} and Erb et al. (2008, for galaxies at $z
\sim$ 2). However, at fixed $\mu_{\rm gas}$ satellites are
typically metal richer than centrals, by 0.06 dex at $\mu_{\rm gas}
\simeq$ 0.1 to 0.12 dex at $\mu_{\rm gas} \simeq$ 0.6. 
This trend likely arises from the anti-correlation between $\mu_{\rm gas}$
and $M_{\star}$, combined with the fact that, at fixed $M_{\star}$, satellites
are metal richer than centrals.
We conclude that the difference in gas phase metallicity
  between centrals and satellites is not associated with a difference
  in (indirectly inferred) gas mass fraction, as would have been
expected in a simple closed box model.
 
\begin{figure*}
\includegraphics[width=130mm]{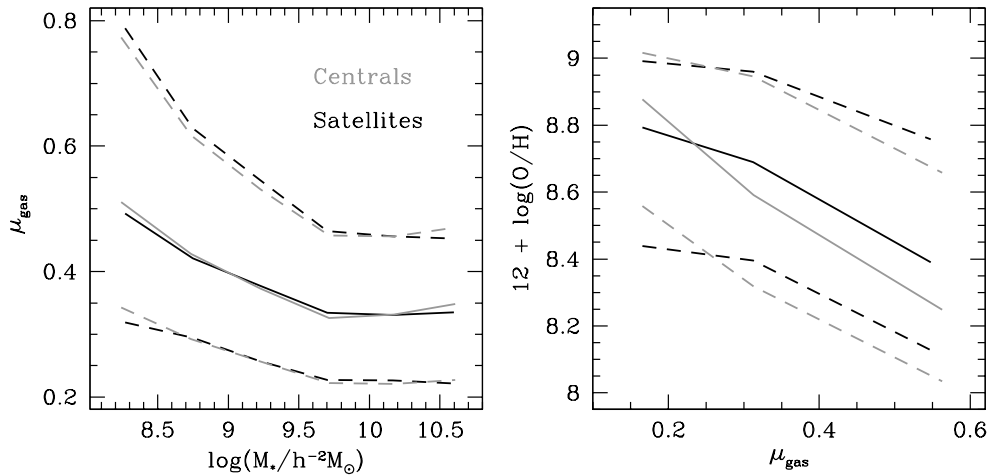}
\caption{{\it Left:} the gas mass fraction in the fibre as a function
  of stellar mass for centrals (grey) and satellites (black). The
  solid lines show the median $\mu_{\rm gas}$ while the dashed lines
  the 16th and 84th percentiles in different bins of stellar mass.
  {\it Right:} the dependence of gas-phase metallicity on $\mu_{\rm
    gas}$ for central (grey) and satellite (black) galaxies.}
\end{figure*}

\subsection{Star formation rates}
The results in the previous two subsections show that the enhanced
gas-phase metallicities of satellite galaxies are not associated with
any enhancement or suppression of the stellar surface mass density or
the gas mass fractions. We now test whether there is any correlation
with the star formation rates.

Fig. 13 shows the specific star formation rate in the fibre
[sSFR$_{\rm fib}$ = log(SFR$_{\rm fib}/m_{\rm fib}$)] and the specific
global star formation rate [sSFR$_{\rm glo}$ = log(SFR$_{\rm
  glo}/M_{\star}$)] as a function of stellar mass for centrals (grey)
and satellites (black) in sample $G$. The sSFR$_{\rm fib}$ of
satellites is higher by only 0.02 dex than that of equally massive
centrals in the range 9 $<$ log($M_{\star}/h^{-2}$M$_{\odot}) <$
10.5. The KS two sample test applied to the cumulative distributions
in sSFR$_{\rm fib}$ of centrals and satellites in different bins of
stellar mass indicates that such a difference is not statistically
significant.\footnote{The same result is obtained when central and
satellite galaxies are compared in their sSFR$_{\rm fib}$ defined as
log(SFR$_{\rm fib}/M_{\star}$).} We checked that the differences
in gas-phase metallicity between satellites and centrals do not 
disappear when galaxies are matched in sSFR$_{\rm fib}$ and stellar
mass.

\begin{figure*}
\includegraphics[width=130mm]{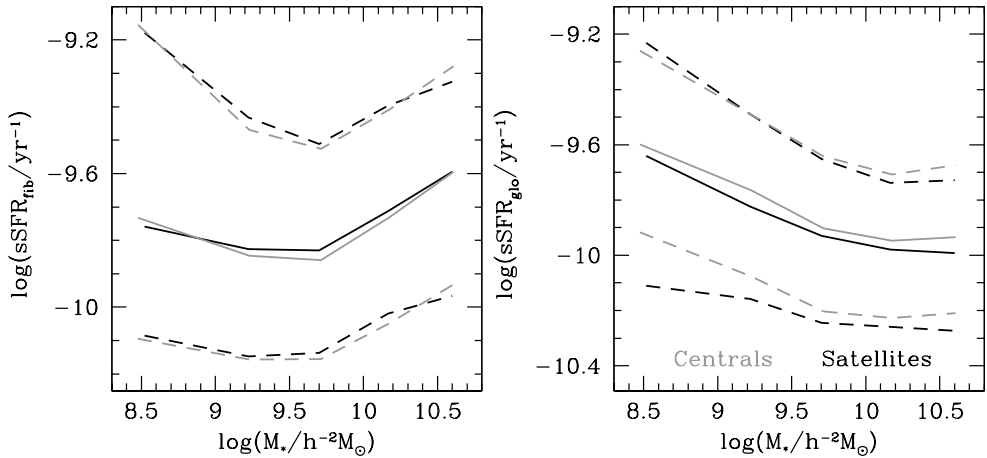}
\caption{The dependence of the specific star formation rate in the
  fibre (sSFR$_{\rm fib}$) and the specific global star formation rate
  (sSFR$_{\rm glo}$) as a function of stellar mass for centrals (grey)
  and satellites (black) in sample $G$. The solid lines indicate the
  median specific SFR and the dashed lines the 16th and 84th
  percentiles of the distribution in specific SFR in different bins of
  stellar mass.}
\end{figure*}

Interestingly, the median sSFR$_{\rm glo}$ of satellites is
systematically lower than that of centrals by 0.03 dex on
average at any stellar mass. The KS two sample test run on the
cumulative distributions in sSFR$_{\rm glo}$ of centrals and
satellites in different bins in $M_{\star}$ indicates that this
difference is statistically robust at $>$95$\%$ confidence level, at
any stellar mass.  The fact that satellites have lower global star
formation rates than centrals of the same stellar mass is consistent
with numerous other studies (e.g., Weinmann et al. 2006, 2009; Kimm et
al. 2007; Cooper et al. 2007; van den Bosch et al. 2008a,b; Pasquali et
al. 2010; Peng et al. 2010; Wetzel, Tinker \& Conroy 2011), and
indicates that certain processes operating on satellite galaxies
causes them to quench their star formation.  The fact that our results
indicate that satellites have lower sSFR$_{\rm glo}$ than centrals, but
are equally active in their central regions (within the fibre),
suggests that this quenching is an outside-in process.

\subsection{Beyond a closed box}

The results in the previous subsections have shown that the differences
in gas-phase metallicities between centrals and satellites cannot be
understood within a simple closed-box model for their chemical
evolution.  This should not come as a surprise, since galaxies are not
expected to be closed boxes. Rather they experience inflows, outflows
and even mass stripping. Interestingly, the efficiency of these
processes is expected to depend strongly on environment, which, as we
argue below, has the potential of explaining the trends identified in
this paper.

\subsubsection{Strangulation}

Central galaxies that reside in host haloes with $M_{\rm h} \lta
10^{12} h^{-1}M_{\odot}$ are expected to accrete cold gas from their
surroundings (`cold flows'), whereas centrals in more massive haloes
are expected to accrete most of their gas through cooling flows from
their hot gaseous haloes (e.g., Birnboim \& Dekel 2003; Kere\v{s} et
al. 2005).  Satellite galaxies, however, are likely to be `deprived'
of these inflows, simply because their gas reservoir, be it hot or
cold, is likely to be stripped by tidal forces and/or ram-pressure
from the hot gaseous atmosphere associated with the host halo. Since inflows
are likely to be metal-poor, this `strangulation' (Larson et al. 1980)
of satellite galaxies is expected to suppress the dilution of their
ISM, resulting in gas-phase metallicities that are higher than those
of central galaxies, in qualitative agreement with the data.
Furthermore, as already discussed in Pasquali et al. (2010),
present-day satellites in more massive haloes were, on average,
accreted earlier (see also De Lucia et al. 2012). Hence, satellites in
more massive haloes have been `deprived of dilution' for a longer time
than satellites of the same stellar mass in less massive haloes, which
might explain why the gas-phase metallicity of satellites increases
with host halo mass (see Fig. 7), at least qualitatively.

However, strangulation is unlikely to be the entire picture.  First of
all, if the differences in gas-phase metallicity (within the fibre)
are due to differences in the amount of dilution, this should probably
show up as differences in the gas mass fraction (again within the
fibre).  As is evident from Fig. 12, such difference is absent,
although we caution that these `gas mass fractions' are indirectly
inferred from the star formation rates (within the fibre). We conclude
that strangulation is likely to play a role in regulating the
gas-phase metallicities of satellite galaxies, but that direct data on
the gas mass fractions of satellite galaxies, combined with detailed
hydrodynamical simulations of strangulation, are needed to test
whether strangulation in itself can explain the various trends
presented in this paper.

\subsubsection{Ram-Pressure Stripping and Galaxy Harassment}

In addition to strangulation,
there are two other mechanisms that may impact gas-phase metallicites
and which operate on satellites but not on centrals: ram-pressure
stripping (e.g., Gunn \& Gott 1972) and galaxy harassment (e.g.,
Farouki \& Shapiro 1981; Moore et al. 1996). We now discuss how
these two processes may cause satellite galaxies to have higher
gas-phase metallicities than centrals of the same stellar mass.

An important mechanism for regulating the gas-phase metallicities of
(the inner regions of) disk galaxies is radial gas flows.  Angular
momentum redistribution within the disk, due to torques from bars,
spirals and/or interacting galaxies can cause the cold gas at large
galactocentric radii to move inwards (cf. Ro\v{s}kar et al. 2008;
Minchev et al. 2012). Since galaxies typically have
metallicity gradients that decline outwards, such radial gas flows
will dilute the gas in the inner regions. Since satellite galaxies are
exposed to stronger tidal forces, and to high-speed impulsive
encounters (i.e., galaxy harassment), they are more likely to
experience such diluting radial flows than central galaxies. If this
were an important mechanism, satellite galaxies would have central
gas-phase metallicities that are lower than central galaxies, opposite
to what is observed. This suggests that, contrary to what we
postulated above, satellite galaxies do {\it not} experience (an
enhancement in) radial gas flows. This could come about if
ram-pressure and/or tidal forces strip the outer, metal-poor gas
before it has the opportunity to migrate inwards. In that case,
satellite galaxies are less likely to experience diluting, radial gas
flows than centrals, which again might explain why the latter have
lower gas-phase metallicities than the former (see also \citet{b30}
and \citet{b32}).

There is ample observational evidence to support the notion that
ram-pressure stripping is an important mechanism, at least in cluster
environments (e.g., Gavazzi et al. 2001; Solanes et al. 2001; Koopmann
\& Kenney 2004; Abramson et al. 2011).  This is also supported by
detailed hydrodynamical simulations (e.g., Abadi et al. 1999; Roediger
\& Hensler 2005; Kronberger et al. 2008; Bekki 2009). However, the
overall efficiency of ram-pressure stripping (i.e.,
what fraction of the gas is stripped and on what time scale) is still
a matter of debate, mainly because the simulations have shown that it
is a strong function of the porosity of the galaxy's ISM (e.g., Quilis
et al. 2000; Tonnesen \& Bryan 2009).  What is well established,
though, is that ram-pressure stripping is more efficient for less
massive galaxies in more massive environments (e.g., Gunn \& Gott
1972; Bekki 2009). This makes the simple prediction that the
suppression of metallicity-dilution in satellite galaxies increases
with decreasing stellar mass and with increasing host halo mass; this
is consistent with the observed trends, at least
qualitatively. Another effect that may contribute here is again the
distribution of infall times; since low mass satellites are typically
accreted earlier than their massive counterparts (e.g., De Lucia et
al. 2012), they have been exposed to the `corrosion' exerted by
their environments for a longer period.  If ram-pressure stripping
indeed causes satellite galaxies to have gas-phase metallicities that
are higher (i.e., less diluted) than those of centrals of the same
stellar mass, then the statistics of infall times would contribute to
explaining the observed trends of gas-phase metallicities of
satellites as a function of $M_{\star}$ and $M_{\rm h}$.

An interesting consequence of ram-pressure stripping, which may have
important consequences for our discussion, is that the gas that is not
stripped away, which is mainly the gas at small galactocentric radii,
may actually be compressed by the ram-pressure, giving rise to
(significantly) enhanced star formation in the disk (e.g., Dressler \&
Gunn 1983; Gavazzi et al. 1995; Kronberger et al. 2008; Kapferer et
al. 2009). Our results that satellite galaxies have the same SFRs in
their central regions (within the fibre) as centrals of the same
stellar mass (see Fig. 13), therefore seems to argue that ram-pressure
stripping is not very efficient. On the other hand, there are clear
indications that the {\it global} SFRs of satellites are suppressed
with respect to those of centrals, which might be an indication that
ram-pressure stripping has started to remove gas from the outer disks.
A ram-pressure stripping induced boost in the SFR will also lead to an
increase in the amount of metals returned to the ISM. However, whether
these metals can be retained or whether they are ejected and/or
stripped is likely to depend on many factors, to the extent that it is
unclear whether this boost in SFR is expected to increase or decrease
the gas-phase metallicities. More detailed studies, in particular of
the HI content of satellites versus centrals, will be needed to
investigate the potential impact of ram-pressure stripping for
explaining the gas-phase metallicities of satellite galaxies.

\subsubsection{Galactic Wind Confinement}

Another environment-dependent process that may potentially play a role
in `regulating' the gas phase metallicities of satellite galaxies is
galactic wind confinement. When a galaxy, central or satellite, is
located within a dark matter halo that has a hot gaseous atmosphere,
the associated pressure may be able to inhibit a potential
supernova-driven galactic wind from escaping the galaxy. Since
galactic winds are typically metal-enhanced (e.g., Mac-Low \& Ferrara
1999; Dalcanton 2007), systems in which the winds can escape are
expected to have lower gas-phase metallicities than systems in which
the wind material is confined to be recycled within the ISM of the
galaxy itself. Since satellite galaxies reside in more massive host
haloes than central galaxies of the same stellar mass, galactic wind
confinement might explain why the former have higher oxygen abundances
than the latter.

This mechanism also nicely explains the observed trends with both
stellar mass and halo mass, at least qualitatively. After all, the hot
gaseous atmosphere in more massive haloes has higher pressure, and
is therefore better able to confine winds. This might explain why the
difference in gas-phase metallicities between centrals and satellites
is larger for satellites in more massive host haloes. In addition,
since winds are expected to be more efficient in less massive haloes,
which host less massive centrals, the potential impact of winds is
more important for less massive galaxies, providing a qualitative
explanation for the scalings with stellar mass. The recent 
simulations developped by Bah\'e et al. (2012) indicate that ram
pressure is larger than the confinement pressure independently
of halo mass. Only a small fraction of the simulated galaxies 
($\sim$16$\%$) appears to be confinement dominated, but this occurs
when they have already gone through the first pericentre passage,
where ram pressure is highest, and have hence lost most of their
hot gas.

\section{Conclusions}

We used the groups catalogue of \citet{b46} together with the DR4
catalogue of stellar ages and metallicities of \citet{b62}, the DR4
compilation of gas-phase metallicities of \citet{b7} and the DR7
catalogue of star formation rates of \citet{b8}, to study the
gas-phase metallicity of central and satellite galaxies and its
dependence on their stellar mass and the dark matter mass of their
host environments. We find that:

\par\noindent

{\it i)} satellites have on average higher gas-phase metallicities 
than central galaxies of the same $M_{\star}$. The magnitude of
this differences depends on stellar mass, increasing from 0.004 dex at
log($M_{\star}/h^{-2}$M$_{\odot}) \simeq$ 10.75 to 0.06 dex at
log($M_{\star}/h^{-2}$M$_{\odot}) \simeq$ 8.25.

\par\noindent

{\it ii)} Within the same host halo, satellites have lower
  gas-phase metallicities than their central galaxies by $\sim$0.5 dex
  on average. This simply reflects that, by definition, central
  galaxies are the most massive galaxies in their halos, combined wih
  the fact that gas-phase metallicity increases with stellar mass.

\par\noindent

{\it iii)} At fixed $M_{\star}$, the gas-phase metallicity of
satellites increases with halo mass; this trend is more pronounced for
less massive galaxies. Low mass satellites with $10^{8.5}
  h^{-2} M_{\odot} \leq M_{\star} \leq 10^{9}h^{-2}M_{\odot}$ have
  average oxygen abundances that increase by $\sim$0.15 dex between
  $M_{\rm h} = 10^{11}h^{-1}$M$_{\odot}$ and $M_{\rm h} =
  10^{14}h^{-1}$M$_{\odot}$.

\par\noindent

{\it iv)} Satellite galaxies have higher gas-phase metallicities
  than central galaxies with the same stellar metallicities. The
  magnitude of this difference increases with decreasing metallicity.

\par 

We have contrasted the gas-phase metallicities of centrals
and satellites with other galaxy properties, such as stellar age,
surface mass density, gas mass fraction and specific star formation
rate, in order to identify the mechanism(s) responsible for the
observational results listed above. Based on these comparisons
we draw the following conclusions:

\begin{itemize}

\item Star-forming centrals and satellites have very similar
    stellar metallicities at fixed $M_{\star}$. This basically rules
    out that the difference in gas-phase metallicities between
    centrals and satellites of the same stellar mass is a consequence
    of stellar mass stripping. Interestingly, this mechanism was used
    by Pasquali et al (2010) in order to explain a similar difference
    between centrals and satellites but in {\it stellar}
    metallicity for mostly quiescent galaxies. 
    The fact that this same mechanism cannot explain the
    gas-phase metallicities is also immediately evident from the fact
    that satellites have higher gas-phase metallicities than centrals
    {\it of the same stellar metallicity} (see Fig. 10).  Rather,
    these results suggest that the differences in 12 $+$ log(O/H) have
    their origin in something that is specific to the gas.

\item Star forming satellites and centrals show very similar
    surface mass densities and gas mass fractions (both measured
    within the fibre) at fixed $M_{\star}$. Consequently, their
    observed offset in gas-phase metallicity remains the same when
    they are matched in surface mass density or gas mass fraction,
    rather than stellar mass. Since these quantities are an indicator
    of the galaxy's past, integrated star formation history (SFH)
    within the fibre we conclude that the differences in 12 $+$
    log(O/H) between centrals and satellites (which are also measured
    within the fibre) are not a consequence of centrals and satellites
    having experienced different closed-box star formation
    histories. Rather, we argue that the mechanism(s) responsible for
    this offset must be associated with differences in their inflow
    and/or outflow histories.

\item Despite their difference in gas-phase metallicities,
    centrals and satellites of the same stellar mass have present-day
    star formation rates, measured within the fibre, that are
    indistinguishable. However, when corrections are made to include
    the star formation occuring on larger scales, outside the
    fibre, we find that satellites have slightly reduced global SFRs
    compared to centrals of the same $M_{\star}$. In addition, we find
    that star forming satellite galaxies are characterized by somewhat
    older stellar ages than equally massive star forming
    centrals. These results suggest that the satellite galaxies have
    started an outside-in process of star formation quenching. We
    emphasize that the overall impact of quenching is small in our
    sample, which is a consequence of the fact that our sample is
    biased towards star forming galaxies, for which gas-phase
    metallicities can be measured from their emission lines.

\end{itemize}

  We argue that there are three mechanisms that operate on satellite
  galaxies and that can potentially explain why they have gas-phase
  metallicities that are higher than those of central galaxies of the
  same stellar mass: (i) strangulation, which inhibits the inflow of
  metal-poor gas, (ii) ram-pressure stripping of the outer, metal-poor
  gas-disk, thereby preventing dilution of the central gas due to
  radial gas flows promoted by torques, and (iii) pressure confinement 
  of metal-enriched outflows. As we have discussed in the
  text, each of these three mechanisms naturally explains why the
  difference in gas-phase metallicity between centrals and satellites
  increases with decreasing stellar mass and with increasing host halo
  mass, at least qualitatively. More work needs to be done in
  order to quantitatively discriminate between these environmental 
  processes. For example, detailed simulations of ram
  pressure stripping and galactic wind confinement, which include also
  star formation and stellar/AGN feedback, are required in order to 
  predict the effect of these mechanisms on the observed stellar and 
  gas-phase metallicities of satellites.
  From an observational point of view, determining the star formation history
  of satellites of different stellar mass and in different environments
  will constrain the time-scale over which strangulation/ram pressure stripping
  quench star formation, while a direct measurement of the (central and total) 
  gas mass of satellites will test the efficiency of strangulation and ram pressure 
  stripping.

\section*{Acknowledgments}

We would like to thank Christy Tremonti for helpful discussions.
\par\noindent
Funding for the SDSS and SDSS-II has been provided by the Alfred P. Sloan Foundation, the Participating Institutions, the National Science Foundation, the U.S. Department of Energy, the National Aeronautics and Space Administration, the Japanese Monbukagakusho, the Max Planck Society, and the Higher Education Funding Council for England. 
The SDSS Web Site is http://www.sdss.org/.
The SDSS is managed by the Astrophysical Research Consortium for the Participating Institutions. The Participating Institutions are the American Museum of Natural History, Astrophysical Institute Potsdam, University of Basel, University of Cambridge, Case Western Reserve University, University of Chicago, Drexel University, Fermilab, the Institute for Advanced Study, the Japan Participation Group, Johns Hopkins University, the Joint Institute for Nuclear Astrophysics, the Kavli Institute for Particle Astrophysics and Cosmology, the Korean Scientist Group, the Chinese Academy of Sciences (LAMOST), Los Alamos National Laboratory, the Max-Planck-Institute for Astronomy (MPIA), the Max-Planck-Institute for Astrophysics (MPA), New Mexico State University, Ohio State University, University of Pittsburgh, University of Portsmouth, Princeton University, the United States Naval Observatory, and the University of Washington.

\end{document}